# Identical Response of Insulators to Irradiations by Swift Heavy Ions: Application to Experiments on Gd$_2$Ti$_2$O$_7$

G. Szenes

Department of Materials Physics, Eötvös University,

P.O. Box 32, H-1518 Budapest, Hungary

E-mail: szenes.gyorgy@ttk.elte.hu

Abstract

Track radii $R_e$ induced by swift heavy ions are studied. The experimental track data are analyzed in $Gd_2Ti_2O_7$ pyrochlore in the range $S_e$=6-54 keV/nm ($S_e$ - electronic stopping power) including irradiations by low and high velocity ions. The Analytical Thermal Spike Model (ATSM) is applied whose main features are reviewed. The track data of $Gd_2Ti_2O_7$ exhibit scaling features which mean simple quantitative relationships with track sizes of other insulators controlled only by a single materials parameter (MP), the melting temperature $T_m$ The $R_e^2$-$S_e$ track evolution curve is described in the whole range of $S_e$ without the application of any individual fitting parameter and MPs apart $T_m$; the threshold for track formation $S_{et}$=6.1/14.4 keV/nm for E<2 MeV/nucleon and E>8 MeV/nucleon, respectively. The unique role of $T_m$ is a highly important limitation for the mechanism of track formation. The practical consequences of the results are discussed with respect of the estimation of irradiation parameters.

# 1. Introduction

The interaction of energetic ions with solids is being investigated with a continuous interest which is demonstrated by the numerous theoretical and experimental studies that have been published in recent years. Intensive research activity is going on materials with pyrochlore structure because of their possible applications for the immobilization and disposal of actinides produced in nuclear plants [1,2]. On the other hand, these materials are very interesting from a scientific point of view, as well, as they exhibit a high chemical and structural flexibility, and their structure has a high radiation tolerance [3].

The stability of these solids under irradiation is a key problem for applications. The irradiation with high energy heavy ions provides a suitable simulation method for the investigation with respect of the long-term effect of actinides. Systematic studies would be required for making reliable predictions. However, in spite of the considerable number of publications only a single pyrochlore $Gd_2Ti_2O_7$ has been studied systematically in a broad range of ion energies E and electronic stopping power $S_e$ [4,5,6] including track formation, as well.

In this paper, quantitative relationships observed in different insulators between the $R_e^2$-$S_e$ track evolution curves ($R_e$ – track radius) are taken into account. These were ignored in previous analyses. Based on such observations, recently, there is a considerable progress in the understanding of track formation induced by electronic excitation [7]. Concerning the effects of materials parameters (MPs) it has been shown that track formation is controlled only by the melting temperature $T_m$ in a broad range of ion energies [7]. Besides practical aspects, this is of high theoretical interest as $T_m$ is an equilibrium parameter, nevertheless, it controls a process under spike conditions. In this paper, the validity of the model is checked for $Gd_2Ti_2O_7$.

The ion-induced temperature distribution $\Delta T(r,t)$ denotes the increase of the local temperature T over the irradiation temperature $T_{ir}$. It is one of the most important information characterizing the response of solids to the impact of the projectiles. Therefore, the reliability of theories depends on a great extent on the accuracy of the calculation of the temperature. The direct comparison with experiments is rather limited in this case and new possibilities are considered.

An emphasis is made throughout the paper on the demonstration of some common features of track formation in various solids. In this respect, the initial value of the track radii $R_{in}$ have an exceptional importance as various subsequent processes may destroy the existing relationship between them. In this paper, the track evolution is described in $Gd_2Ti_2O_7$ at low and high ion velocities, by applying the Analytical Thermal Spike Model (ATSM) [8]. The results in $Gd_2Ti_2O_7$ and other insulators are compared and scaling features are demonstrated.

## 2. Theoretical background

The most widely studied irradiation effect is the formation of ion-induced tracks. It was first observed in 1959 in mica after exposing to uranium fission fragments [9]. In the following more than 60 years, the theoretical efforts were concentrated in finding the right mechanism linking the physical properties of the actual target and the irradiation parameters with the track size. However, it was shown later that this could not lead to satisfactory result as there was found additionally a clear quantitative relationship between track radii induced in different solids by different ions of different energies [10].

The above results have been incorporated into the ATSM published first in 1995 [8]. In the ATSM, it is assumed that the Gaussian function is a good approximation for the ion-induced temperature distribution. Equations are derived for the threshold value $S_{et}$ for track formation and for the track evolution function $R_e^2(S_e)$ [8]. When ATSM was applied to track forming insulators, a simple relationship was observed between track radii measured in different insulators [7]. According to this relation track radii do not depend on MPs other than $T_m$. Track radii are controlled by the ion-induced temperature distribution. When a quantitative relation exists between track radii, there must be a quantitative relation between the induced temperatures, either.

A systematic study of track formation led to revealing that at the moment of maximum peak temperature $T_p$ (t = 0) the ion-induced temperature increase $\Delta T(r,0)$ is identical in various insulators for $<s_e>$=constant [7]. This universal-type temperature distribution is given by

$$\Theta(r) = \frac{f<s_e>}{3\pi k w^2} \exp\{-r^2/w^2\} \qquad (1)$$

where $<s_e> = S_e/N$ (N- number density of atoms), f is the efficiency, k is the Boltzmann constant and w=4.5 nm for insulators. In insulators, low velocity ions induce larger tracks than high velocity ions at $S_e$=constant and this is called velocity effect (VE) [11]. In ATSM, the efficiency f is responsible for VE and f≈0.17 is valid for E>8 MeV/nucleon (HI) and it may change to f≈0.4 [7] in the range E<2 MeV/nucleon (LO). Presently, there is no sufficient experimental information for irradiations within the range 8 MeV/nucleon>E>2 MeV/nucleon and the shape of the transition of the efficiency between the LO and HI values is not known reliably in this range.

The presence of the efficiency f in Equation (1) is the demonstration that the deposited electronic energy is not completely transferred to thermal energy. Recently, this has been confirmed by direct experiments on $Y_3Fe_5O_{12}$ [12]. The comparison of Equation (1) with the experiments in insulators leads to the conclusion that the spatial localization of the energy deposition in the lattice is characterized by the same value w=4.5 nm whatever is E. The term MP is used in the sense of an individual materials property. Therefore, the common parameters f and w in Equation (1) are not considered as an MP in insulators, though they are obviously parameters that are related to materials. In semiconductors, however, w is an MP.

Recently, the uniform temperature distribution described by Equation (1) was checked in the analysis of experimental data on electronic sputtering. Systematic data of 10 solids (amorphizable and non-amorphizable insulators and semiconductors) were analyzed assuming a thermal activation mechanism [13]. Excellent agreement was found with the experiments for all solids in a broad range of $S_e$. The results confirm the validity of Equation (1) when it claims that the ion-induced temperature does not depend on MPs. It was also found that w=constant in a broad range of induced temperature.

Taking into consideration the high complexity of the processes (extremely high local charges, temperatures, stresses induced in a narrow space within a very short time) the uniform behavior is highly unexpected.

Actually, ATSM predicts that the track evolution can be described by the equations [8]

$$R_e^2 = w^2 \ln(S_e/S_{et}), \qquad \text{for } S_e < 2.7 S_{et} \qquad (2)$$

$$R_e^2 = \frac{w^2 S_e}{2.7 S_{et}}, \qquad \text{for } S_e > 2.7 S_{et} \qquad (3)$$

$$S_{et} = \frac{\pi \rho c (T_m - T_{ir}) w^2}{f}, \qquad (4)$$

where $\rho$ and c are the density and specific heat. Often the $\rho c=3Nk$ approximation is used in the equations. The same equations can be derived from Equation (1) as well. In the analyses according to ATSM, a logarithmic function is fitted to the experimental data in the range $S_e<2.7S_{et}$ (see Equation (2)) and w and f are the two parameters of the model derived from this fit. In principle w and f are fitting parameters, but due to specific features of this effect they have the same values for various insulators.

When applying ATSM the above procedure cannot be avoided. When the value of $S_{et}$ provided by another source is used in the analysis this necessarily modifies the derived values of the w and f parameters and in this case Equations (2-4) of ATSM do not describe the track evolution any more. In a review of ATSM by Dufour and Toulemonde, such improper action was made both in the LO and HI ranges [14] and the equations with the derived parameters were applied to track data measured in $Y_3Fe_5O_{12}$. Earlier the same data had been already analyzed by using ATSM correctly [15]. The comparison of the results of the two analyses with the original experimental data is very instructive. While the corresponding plot in Ref.[15] show an excellent agreement with the data, the track sizes calculated in Ref.[14] do not even remind of the original ones. There are also several similar elementary problems with this review in Ref.[14].

It is important that Equations (2-4) are valid for the initial radius $R_{in}$ of the amorphous cylindrical volume formed in the thermal spike. In agreement with the identical temperature distribution expressed by Equation (1) there is a simple unambiguous quantitative relationship between these track sizes and also between the $R_e^2$-$<s_e>$ track evolution curves measured in different insulators. This has been proven for a number of insulators in which $R_{in}$ was measured [7,16].

However, in many experiments $R_{in}$ cannot be estimated because of the fast processes of recrystallization or phase transformation. By now, the validity of Equation (1) has been justified for about 20 insulators. As originally each one has been selected randomly for the experiment, it is reasonable assuming that the validity extends to a much higher number of insulators including those where $R_{in}$ cannot be measured. Actually, the validity of the composition independent, uniform temperature distribution (Equation (1)) must not depend on whether $R_{in}$ can be measured by the applied experimental techniques, or cannot. It is not obvious, what could be the reason that swift heavy ions induce $\Delta T(r,0)$ according to Equation (1) in one group of insulators, where $R_{in}$ can be measured while in another group of insulators where $R_{in}$ cannot be measured, the same ions would induce a totally different temperature distribution depending on many MPs. Much more systematic track data is necessary for the final solution of this problem.

The problems with any theory are revealed when the predictions are compared with the experiments. In this paper such comparison is extended to irradiation experiments on $Gd_2Ti_2O_7$ and it is shown that the predictions of ATSM are in good agreement with the experiments in a broad range of E and $S_e$. This analysis diverges from any previous one on this solid as the track evolution is described without using any individual fitting parameters and MPs – except $T_m$. This is a rather severe condition when the accuracy and the validity of a model is checked.

### 3. Experimental data and their analysis

In Ref.5 a detailed experimental and theoretical study is published on track formation induced by swift heavy ions in $Gd_2Ti_2O_7$ pyrochlore. The samples were irradiated by various ions of 11.1 MeV/nucleon initial energy and transmission electron microscopy (TEM), Raman spectroscopy and X-ray diffraction (XRD) were used for the study of the ion-induced structures. Bright-field and High- resolution TEM methods were also used. The samples were made either by crushing the irradiated specimens or by conventional TEM-specimen preparation, including polishing and ion milling. The original depth in the irradiated sample is unknown for either type of TEM specimens.

It was found that radii measured by TEM correspond to tracks consisting of an amorphous core and a disordered defect-fluorite shell. Whereas radii derived from the XRD maxima are related only to the amorphous track core. Consequently, track radii derived from TEM experiments always exceeded the sizes obtained by XRD method at the same values of $S_e$ [5].

Following other studies [5], only TEM results were used in this analysis. It is a reasonable decision as there is no individual adjustable parameter in this model. Therefore, it is more sensitive to the uncertainties of the experimental data than other models. For the same reason, there is always an endeavor to avoid including those data in the analysis for which the parameters of the model ($T_m$, f, N) or the experimental conditions (E, $S_e$) are not well defined.

Similarly, those track data could not be used when inhomogeneous track structures were reported and there was no information about the initial track size. A practical solution of the problem is when the measurement of the total track radius is possible as it is close to the initial amorphous size with $R_{in}$. This was done in Ref.[5] for $Gd_2Ti_2O_7$ providing a possibility for the application of ATSM.

A further critical point is that there are no suitable systematic track measurements on pyrochlores other than $Gd_2Ti_2O_7$. Obviously, it is very disadvantageous for the application of any model, but especially those without individual adjustable parameters when only a single track size is known. The uncertainty of the derived parameters is considerably increased in such cases. This was the reason that some experimental data had to be ignored.

Though the above limitation reduces the available database, however, the unambiguous demonstration of a close quantitative relationship between track radii and Equation (1) would not be possible without such limitations.

In the experiment in Ref.[5], the effect of the ion velocity changes in the 40 μm thick samples as the ion energy varies in the range 11.1 MeV/nucleon<E<2 MeV/nucleon along the trajectory [5]. While $S_e$≈constant along the trajectory, track radii change by more than 50% because of the VE. Different values of $R_e$ could be obtained in this experiment depending on whether the original position of an actual TEM sample was closer to the front side (E=11.1 MeV/nucleon, HI condition) or to the back side (E=2 MeV/nucleon, LO condition) in the thick specimen. Thus, the experimental method led to an additional minimum ±25% error in the track diameters in Ref.[5]. According to the above considerations, the VE is a source of considerable uncertainty for track studies in this type of irradiation experiments.

In another experiment, $Gd_2Ti_2O_7$ was irradiated by 120 MeV U ions [4]. While the irradiations were performed with HI ions in Ref.[5], all data in Ref.[4] belonged to the low velocity range with E<2 MeV/nucleon. Beside monoatomic irradiation by U ions, $Gd_2Ti_2O_7$ samples were also irradiated by 30 MeV C60 ions that completed the LO data [16]. The experiments in Refs.[4,5,16] were performed in a broad range of $S_e$ (5-54 keV/nm) and E (0.04-11 MeV/nucleon).

In Figure 1, track radii induced by HI [5] and LO projectiles [4,16] in $Gd_2Ti_2O_7$ are shown in a normalized plot where ρc=3Nk is used. The internal structure of tracks is ignored here and the experimental $R_e$ radii with amorphous core+defect-fluorite shell structure are used as discussed above. In the figure, the solid lines are the predictions according to Equation (2) with w=4.5 nm f=0.17 for HI irradiations (E>8 MeV/nucleon) and f=0.4 for LO irradiations (E<2 MeV/nucleon). Equation (2) is valid for $S_e$<2.7$S_{et}$ and predicts an $R_e^2 \propto w^2 \ln S_e$ dependence in agreement with the experiments. When $S_e$ is higher, $R_e$ cannot be directly obtained from Equation (2) as the maximum radius of the melt is reached at t>0 during the cooling phase [8]. In this case, Equations (3,4) are combined leading to

$$R_e^2 = \frac{1}{2.7} \frac{fS_e}{3\pi kN(T_m - T_{ir})} \qquad \text{for } \frac{S_e}{S_{et}} > 2.7. \qquad (5)$$

It is remarkable that Equation (5) does not depend on w. The same is true for Equation (3) as Equation (4) is always valid in ATSM. This was not realized in the review in Ref.[14] when the value of w was estimated using exclusively Equation (3) and ignoring Equation (4). This led to a considerable error.

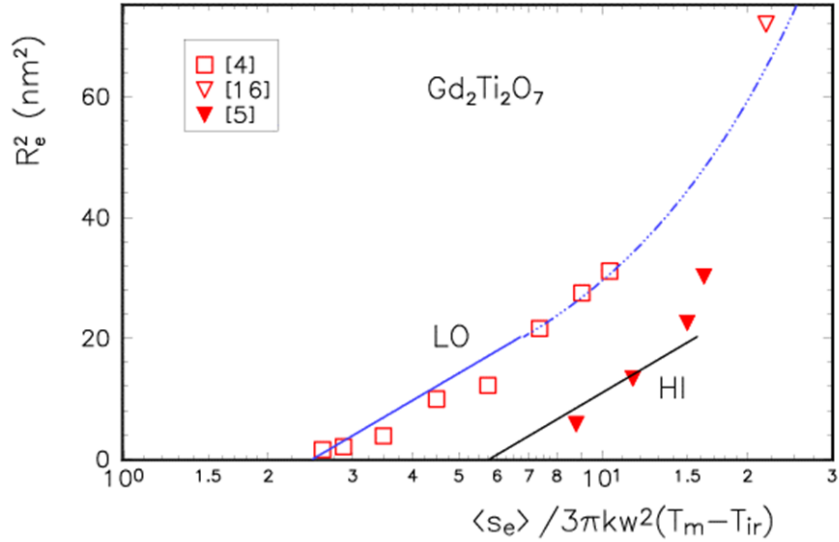

*Figure 1. Variation of the track size in $Gd_2Ti_2O_7$ [4,5,16] with the atomic stopping power $<s_e>$; $R_e$ and $k$ are the track radius and the Boltzmann constant, $w=4.5$ nm and $T_m=2093$ K [6,16] is the melting and $T_{ir}$ the irradiation temperatures. The irradiation by high velocity projectiles is denoted by full symbols. The solid and dashed lines are drawn according to Equation (2) and Equation (5), respectively, without fitting parameter.*

Equation (5) is valid at high values of $S_e$ and the dashed curve in Figure 1 was calculated applying it. It is emphasized the good agreement between the data ($\Delta R_e/R_e \approx 5\%$, $\Delta S_e/S_e \approx 5\%$) and the theoretical predictions for irradiation by LO ions including C60 irradiation when $S_e$ and E are known accurately [4,16]. The agreement was achieved when $S_e$ varied by an order of magnitude and no individual fitting parameters were used. This verifies that Equations (2-5) provide an accurate description of the track evolution in $Gd_2Ti_2O_7$ that depends only on a single MP - $T_m$, and $\Delta T(r,0)$ is correctly described by Equation (1) with Gaussian width w that does not change with E.

The comparison of the track data from Ref.[5] with the predicted HI curve shows lower than expected uncertainty in the positions of the TEM samples in the thick specimen. The figure suggests that except the largest track at $S_e=40.1$ keV/nm, the original positions of the TEM samples were occasionally in that half of the irradiated specimen which was close to the impact of the projectile. The agreement of the predicted curve with the experimental data in the HI range is satisfactory when the possible consequences of the thick samples in Ref.[5] (leading to

±50% uncertainty for $R_e^2$) are taken into account. As the random selection of the TEM samples along the thickness in Ref.[5] and the existence of VE in $Gd_2Ti_2O_7$ are without doubts, the above explanation seems to be reasonable. Thus this high error is simply the consequence of VE and it is not related to the estimates using ATSM.

When Equation (1) is applied in Figure 1, $fS_{et}=E_T$ is obtained for $R_e=0$ where $E_T=3\pi Nkw^2(T_m-T_{ir})$ is the energy necessary for increasing the maximum temperature of the spike to $T_m$. Estimates for the thresholds are $S_{et}$=14.4 keV/nm (HI, f=0.17, $T_m$=2093 K [6,16] $1/E_T$=0.408 nm/keV) and $S_{et}$=6.13 keV/nm (LO, f=0.4, $1/E_T$=0.408 nm/keV).

In Figure 2, the normalized threshold values for track formation $<s_{et}>$ are shown for various insulators. The comparison of the present results with the data in other insulators in Figure 2 shows that these estimates for $Gd_2Ti_2O_7$ nicely fit to the lines in the figure and verify their reliability. According to Equation (4) the slope of the line is $m=\pi w^2/f$ in Figure 2 (w=4.5 nm, f=0.17/0.4), which is in excellent quantitative agreement (within 5%) with the experiments.

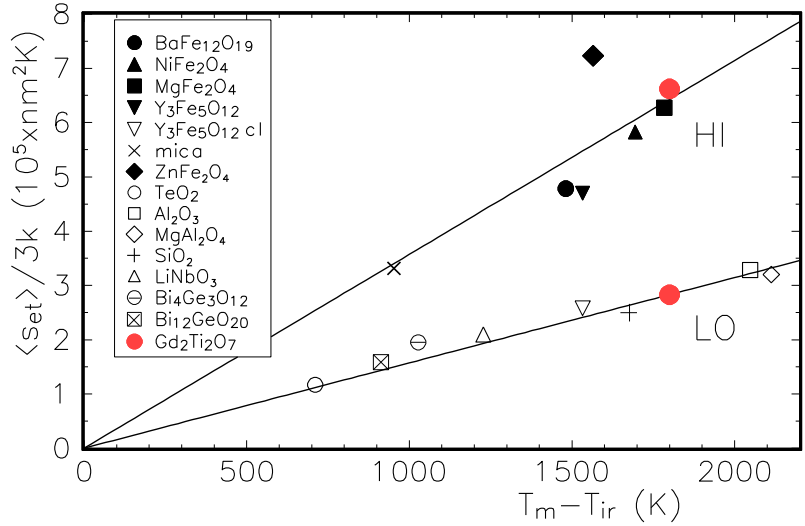

*Figure 2. Variation of the normalized threshold electronic stopping power $<s_{et}>$ at low and high ion velocities; $T_m$ and $T_{ir}$ are the melting and the irradiation temperatures, k is the Boltzmann constant. The values for $Gd_2Ti_2O_7$ are present results.*

It is an interesting feature of the plot in Figure 2, that it can be used for estimates of the value of $S_{et}$ for transformed tracks as well since they cannot skip this stage of formation [15].

The relationship between track sizes is demonstrated by another plot. When Equations (2,4) are combined the result predicts that the $R_e^2$-$<s_e>$ track evolution curves follow the same lines in an $R_e^2$ - $f<s_e>/3\pi kw^2(T_m-T_{ir})$ plot [7]. In Figure 3, such plot is shown for tracks induced by HI ions (SrFe$_{12}$O$_{19}$ [17], MgFe$_2$O$_4$ [18], NiFe$_2$O$_4$ [19] ZrSiO$_4$ [20], Y$_2$O$_3$ [21]), together with data induced by LO ions (mica [7], Y$_3$Al$_5$O$_{12}$ [22], TeO$_2$ [23], Y$_3$Fe$_5$O$_{12}$ [11,24], Al$_2$O$_3$ [15], KTiOPO$_4$ [25], LiNbO3 [26]). The lines cross the axis at $<s_e>/3\pi kw^2(T_m-T_{ir})=1/f$ providing f≈0.4 (LO) and f≈0.17 (HI). The figure proves that track formation proceeds identically in these solids including Gd$_2$Ti$_2$O$_7$. This means that similarly to the plot in Figure 1, the same description of the track evolution is valid for solids in Figure 3, as well.

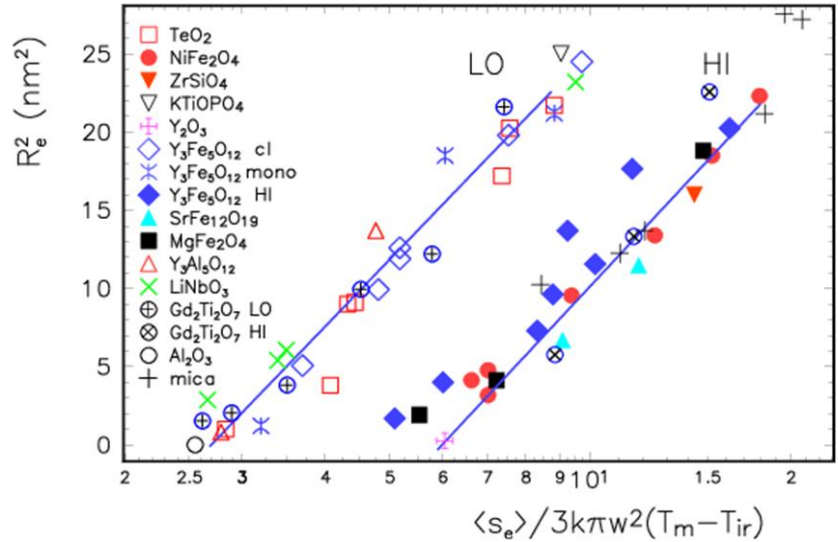

*Figure 3 Variation of the track size with the atomic stopping power $<s_e>$; $R_e$ and $k$ are the track radius and the Boltzmann constant, $w=4.5$ nm and $T_m$ and $T_{ir}$ are the melting and the irradiation temperatures. The lines are drawn according to Equations (2,4). The tracks were induced by projectiles with $E < 2$ MeV/nucleon (LO) and $E > 8$ MeV/nucleon (HI) (for references see the text).*

It is noted that in Figure 3 the track sizes with highest deviations were measured in Y$_3$Fe$_5$O$_{12}$ HI samples having thicknesses typically over 60 µm in the experiments. These high deviations are attributed partly to VE along the track length leading to the formation of larger tracks approaching the backside of the samples. VE had similar effect also in Gd$_2$Ti$_2$O$_5$ as 40 µm thick samples were irradiated in Ref.[5].

An unexpected relationship between track radii measured in different insulators is demonstrated in Figure 4 where the appropriate $R_e$ values for $Gd_2Ti_2O_7$ are also included. Thus $Gd_2Ti_2O_7$ belongs to that group of insulators whose response to swift ion irradiations has common features.

It is emphasized that the relationship in Figure 4 has been predicted by using ATSM [10]. For deriving this plot, Equations (2,4) were combined and a plot was drawn at $<s_e>$=constant. Evidently, similar plots can be obtained for other values of $<s_e>$ as well. This is an important evidence supporting ATSM.

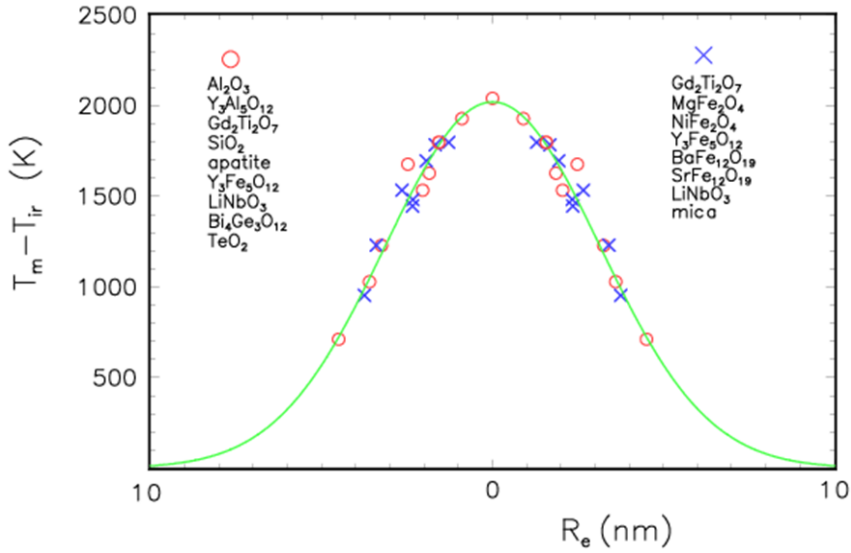

*Figure 4. Variation of track radii $R_e$ with the melting temperature $T_m$ in insulators irradiated by low (o) (E<2 MeV/nucleon) and high velocity (x) (E>8 MeV/nucleon) ions at $<s_e>/3k=S_e/3Nk=3.42x10^5$ nm$^2$ K and $7.5 \times 10^5$ nm$^2$ K, respectively, for track data see [27] (N – number density of atoms, k –Boltzmann constant, $T_{ir}$ – irradiation temperature). In the figure, the solids are listed in the order of decreasing melting temperatures $T_m$. The enveloping curve $\Theta(r)=T_p exp\{-r^2/w^2\}$ is a fit with w=(4.5 ± 0.3) nm and $T_p$=2020 ± 60 K [28].*

Figure 4 is an unambiguous evidence confirming the validity of the thermal mechanism of track formation since $T_m$ is the unique controlling parameter of the process. Actually, it is in agreement with the assumption of the thermal spike models that amorphous tracks are formed at $T=T_m$ by the fast cooling and solidification of the melted cylindrical volume along the trajectory of the projectiles.

It is emphasized that the original experimental data have a primary significance for any theory. In this respect it is highly important that the message of Figure 4 is not modified by any individual adjustable parameter and other MPs with unreliable values under spike conditions.

It is a remarkably important experimental evidence shown by this plot that MPs apart $T_m$ (including the heat of fusion) have no effect on the track size. In the opposite case high deviations from the enveloping curve would appear. This is a serious restriction for any mechanism proposed for the ion induced track formation.

The plot in Figure 4 is a basic element for the understanding of the mechanism of VE. In this respect, it is important, that the widths of the Gaussian curves are identical for LO and HI irradiations and only the thermal energy of the spike varies with E.

The common behavior demonstrated in Figures 2-4 is exhibited by about 20 insulators including $Gd_2Ti_2O_5$. This is a rather high number as these solids have not been specially selected in the original experiments. Taking into consideration the high complexity of the processes (extremely high local charges, temperatures, stresses induced in a narrow space within a very short time) the uniform behavior in Figures 3,4 is highly unexpected.

We note that, previously, track evolution has been already analyzed in $Gd_2Ti_2O_7$ applying the i-TS model [14] requiring an individual adjustable parameter and the knowledge of a considerable number of MPs under spike conditions and assuming a superheating scenario [29]. The advantage of the description given by ATSM seems to be without doubt as it is valid in a broader range of $S_e$ with applying only a single MP without any individual fitting parameter.

## 4. Discussion
### 4.1. The role materials parameters

According to the thermal spike philosophy, the most important information is the accurate description of the ion-induced temperature. ATSM offers a common solution for insulators, while it ignores the heat of fusion L and other MPs. Otherwise, the contribution of L might affect considerably the temperatures. In a recent paper [28] this problem was discussed in details and most of the available systematic experimental track data were analyzed. The results confirmed that L might have only a minor effect on track formation. The present analysis of track evolution in $Gd_2Ti_2O_7$ also refutes the usual argument according to which the good agreement with the experimental track sizes is a sufficient evidence for the validity of a model. In this case it would not be possible that several models could lead to equally good agreement with a given set of data, while only one of the proposed models may be valid. The correct model must provide a complete description of the irradiation effect and explain beyond the track sizes the existing relations between track formation in different solids (e.g. Figures 2-4).

In this paper, the theoretical lines in Figure 1 are drawn in agreement with Equation (1) which does not contain any contribution of L. As $L/E_T$ varies in a broad range in insulators, this large variation ought to lead to considerable shifts of $<s_{et}>$ (crossing of the lines with the X-axis) to higher values in Figure 4. However, no such deviations can be observed in the figure. On the other hand, the good agreement with the experiments in a broad range of $S_e$ in Figure 1 would not be possible if ignoring L were not correct. The smooth curve in Figure 3 is also a strong evidence supporting this conclusion.

The above result concerning L has general consequences but, actually, it leads to important conclusions for $Gd_2Ti_2O_7$, as well. In Ref.[5] $S_{et}$=11.7 keV/nm is derived for HI tracks that corresponds to 4.8 in the X-axis in Figure 4. In the i-TS model, the threshold $S_{et}$ =11.7 keV/nm contains the energy for heating to $T_m$ and the contribution of the latent heat as well [30]. Thus, $T_m$ ought to be reached already at about $S_e \approx 7.7$ keV/nm which corresponds to 7.7x0.408=3.14 on the X-axis in Figure 4.

This is about 50% deviation from the value which is derived based on the common behavior of insulators demonstrated in Figures 2-4. The difference is even higher for LO irradiation as only $S_{et} \approx 3$ keV/nm is estimated in Ref.[16]. Thus, the induced temperature would reach $T_m$ at about $S_e$=2 keV/nm, that is even a much lower value. Consequently, the ion-induced temperatures are much overestimated for $Gd_2Ti_2O_7$ in Ref.[5]. While the experimental data show a close relationship between $S_{et}$ and $T_m$ in Figure 2, $S_{et}$ is related to a much higher superheating temperature, $T_{ps}>T_m$ in Ref.[5]. This is the reason that the temperature is supposed to be raised to $T_m$ at a much lower value of $S_e$ in that model.

A possible origin of the discrepancy might be that the estimates for superheating were made for solids without taking into account the ion-induced transient processes and the highly damaged structure of the targets. These specific features may affect considerably the proceeding of various processes including phase transition.

As the reliable knowledge of the temperature is the basis for any calculation in the spike, therefore, various estimates made for the irradiated $Gd_2Ti_2O_7$ are problematic in Ref.[5]. The same is valid for all those solids whose data are included in Figures 3-4 and similar plots. It is important in this respect that while $Y_3Fe_5O_{12}$ is an emblematic insulator for i-TS, the most popular model denying scaling properties of ion-induced tracks [30], nevertheless, its experimental data fit well to other ones in all those figures which demonstrate the interdependence of tracks. This means that the $R_e^2$-$S_e$ track evolution curve can be calculated easily for $Y_3Fe_5O_{12}$ and other solids in Figures 3,4 using track data of any other insulator in these figures. To do that only $T_m$ values must be known, other MPs are indifferent. This has been demonstrated in Ref.[7] and agreement between the calculated and measured values was found within experimental error. In the same time, up to 10 MPs are required for the description of the same tracks in $Y_3Fe_5O_{12}$ [30] and $Gd_2Ti_2O_7$ [5]. However, it remains an open question whether solutions in Refs.[5,30] can be considered as a complete solution of the problem when the interdependence of these tracks is not even mentioned.

The possibility of the above quite different two solutions is provided by the fact that the width of $\Delta T(r,0)$ can be checked by track measurements only at a single temperature, $T=T_m$. The shape of $\Delta T(r,0)$ cannot be controlled at any other temperature. Thus a great number of various distributions with the same width at $T=T_m$ satisfy this simple condition and any deviation from the real distribution at $T \neq T_m$ is indifferent for the agreement with experiments. However, when a quantitative relationship is valid between track radii in different solids, then the width of $\Delta T(r,0)$ is also determined at several temperatures (see Figure 3) thus its shape can be checked in several points. These considerations may be useful for a correct understanding of an agreement between measured and calculated values.

The above results indicate that the correct estimation of the temperature is highly important and indispensable in cases when a quantitative comparison is possible with experiments. The plots in Figures 2-4 and the accurate prediction of the $R_e^2$-$S_e$ track evolution curve of $Gd_2Ti_2O_7$ are sound evidences confirming the validity of Equation (1). Up to now, the meaning of these figures have not been discussed and explained during the last 15 years using any other model. Thus the readers have no information how the various models comply with the new experimental limitations.

We see that some important features of track formation can be obtained directly from Equation (1) and Figures 2-4. Evidently, the specific features of the amorphous track core + disordered shell with defect-fluorite structure in $Gd_2Ti_2O_7$ cannot be derived from Equation (1). This is a very complex nanostructural problem [5]. Its discussion is beyond the scope of this paper.

## 5. Conclusions

When initial track radii are known in insulators, common scaling features show up and $T_m$ is the only MP which controls the track size. These scaling features include i./ $<s_{et}> \propto (T_m-T_{ir})$ in Figure 2 for the threshold atomic stopping power; ii./ common $R_e^2$-$<s_e>/(T_m-T_{ir})$ track evolution curves in Figure 3; iii./ common $R_e$ - $T_m$-$T_{ir}$ Gaussian curve controlled by $T_m$ in Figure 4 connecting track data measured in various solids for $<s_e>$=constant.

The validity of the above relations is supported by experimental data on ion-induced tracks measured in various insulators including $Y_3Fe_5O_{12}$. This is a serious contradiction as this solid is an emblematic prototype for the i-TS model. The scaling features are also demonstrated in $Gd_2Ti_2O_7$ where the measured $R_e$ total track radii with amorphous core+defect-fluorite shell structure are close to the initial radii. The experimental $R_e^2 – S_e$ values satisfy the equations of ATSM for a very broad spectrum of $S_e$ (5-54 keV/nm) and E (0.04-11 MeV/nucleon) without applying any individual fitting parameter and without MPs except $T_m$. LO and HI tracks in $Gd_2Ti_2O_7$ are included in Figure 4 that demonstrates the interdependence of tracks induced in different insulators when $<s_e>$=constant.

The results on $Gd_2Ti_2O_7$ confirm the validity of Equation (1) claiming that the ion-induced temperature distributions agree quantitatively in numerous insulators for identical values of $<s_e>$. The estimates of the thresholds for $Gd_2Ti_2O_7$ are $S_{et}$=6.13 and 14.4 keV/nm for LO and HI projectiles, respectively, in agreement with the values derived from the scaling properties. The heat of fusion L may have only minor if any effect on the track size. Thus, the assumption of the superheating mechanism is not justified by the experiments. It is a source of a systematic error in the calculation of the ion-induced temperatures that may exceed even 50%.

**Disclosure statement**

The author declare no conflict of interest.